\documentclass[pre,twocolumn]{revtex4}
\usepackage{graphicx}
\usepackage{subfigure}

\newcommand{\product}{\prod}
\newcommand{\figpath}{}

\newcommand{\etal}{\textit{et al.}}

\newcommand{\ie}{i.e.}

\newcommand{\figrefformat}[1]{\ref{#1}}
\newcommand{\figref}[1]{Fig.~\figrefformat{#1}}

\newcommand{\sectionrefformat}[1]{\ref{#1}}
\newcommand{\sectionref}[1]{section~\sectionrefformat{#1}}

\newcommand{\given}{\! \mid \!}
\newcommand{\xbar}{\bar{x}}

\newcommand{\puncspace}{\enspace}
\newcommand{\mathcomma}{\puncspace ,}
\newcommand{\mathperiod}{\puncspace .}

\newcommand{\pa}[1]{\ensuremath{\mathrm{Pa}(#1)}} 

\newcommand{\set}[1]{\ensuremath{ \left \{#1 \right \}}}

\newcommand{\rhoest}{\hat{\rho}}

\newcommand{\meanval}[1]{\ensuremath{\xbar_{#1}}}

\newcommand{\maplinear}[2]{\ensuremath{A^{#1}_{#2}}}
\newcommand{\mapconst}[2]{\ensuremath{B^{#1}_{#2}}}
\newcommand{\mapfunc}[1]{\ensuremath{g\left(#1\right)}}
\newcommand{\act}[2]{\ensuremath{a^{#1}_{#2}}}

\newcommand{\decoder}[2]{\phi^{#1}_{#2}\left(x_{#1}\right)}
\newcommand{\decodeconst}[2]{\xbar^{#1}_{#2}}
\newcommand{\coupling}[2]{\ensuremath{K_{#1#2}}}

\newcommand{\externsignal}[2]{\ensuremath{h^{#1}_{#2}}}
\newcommand{\stablewts}[3]{\ensuremath{S^{#1}_{#2#3}}}
\newcommand{\selfwts}[3]{\ensuremath{T^{#1}_{#2#3}}}

\newcommand{\ffwts}[4]{\ensuremath{U^{#1#2}_{#3#4}}}
\newcommand{\fbwts}[4]{\ensuremath{V^{#2#1}_{#3#4}}}
\newcommand{\indirectwts}[5]{\ensuremath{W^{#1#2#3}_{#4#5}}}

\newcommand{\wtdefwrap}[4]{\maplinear{#1}{#2}{#4}\decodeconst{#1}{#3}}
\newcommand{\varinv}[1]{\frac{1}{\sigma_{#1}^2}}
\newcommand{\couplevar}[2]{\varinv{#1}\coupling{#1}{#2}}

\begin{document}

% Definition of title page:
\title{Neural Networks Processing Mean Values of Random Variables}
\author{M.~J.~Barber}
\email{mjb@uma.pt}
\affiliation{Universidade da Madeira,
    Centro de Ci\^encias Matem\'aticas,
    Campus Universit\'ario da Penteada,
    9000-390 Funchal, 
    Portugal}
\author{J.~W.~Clark}
\affiliation{Department of Physics,
    Washington University,
    Saint Louis, MO 63130}
\author{C.~H.~Anderson}
\affiliation{Department of Anatomy and Neurobiology,
    Washington University School of Medicine,
    Saint Louis, MO 63110
}

\date{\today}    

\begin{abstract}
    We introduce a class of neural networks derived from 
    probabilistic models in the form of Bayesian belief networks.
    By imposing additional assumptions about 
    the nature of the probabilistic models represented
    in the belief networks, 
    we derive  neural 
    networks with standard dynamics
    that require no training to determine the 
    synaptic weights, 
    that can pool multiple sources of evidence, and
    that deal cleanly and consistently with inconsistent
    or contradictory evidence. 
    The presented neural networks capture many properties of
    Bayesian belief networks, providing distributed versions
    of probabilistic models.
\end{abstract}

\maketitle

\section{Introduction}\label{sec:intro}

Strong feedforward, feedback, and lateral connections  exist between 
distinct areas of the cerebral cortex, but such connections are not 
observed in cerebellar, sensory, or motor output circuits.  The 
anatomical structure of the cerebral cortex may facilitate a modular 
approach to solving complex problems \cite{vanessen/etal:1992}, with 
different cortical areas being specialized for different information 
processing tasks.  To permit  a modular strategy of this sort, 
coordinated and efficient routing of information must be maintained 
between modules, which in turn demands extensive connections 
throughout the cortex.

It has been proposed \cite{anderson:1994} that cortical circuits
perform statistical inference, encoding and processing information
about analog variables in the form of probability density functions
(PDFs).  This hypothesis provides a theoretical framework for
understanding diverse results of neurobiological experiments,
and a practical framework  
for the construction of recurrent
neural network models that implement a broad variety of 
information-processing functions 
\cite{barber/clark/anderson:2003a,eliasmith/anderson:1999,%
eliasmith/anderson:2002}.

Probabilistic formulations of neural information processing have been
explored along a number of avenues.  
One of the earliest such analyses 
showed that the original Hopfield neural network implements, in 
effect, Bayesian inference on analog quantities in terms of PDFs 
\cite{anderson/abrahams:1987}.   
%: HOPPs could be mentioned here
As in the present work,
Zemel \etal\ \cite{zemel/etal:1998} have
investigated population coding of probability distributions, but with 
different representations and dynamics
than those we will consider here.  Several extensions of this representation 
scheme have been developed 
\cite{zemel:1999,zemel/dayan:1999,yang/zemel:2000} that 
feature information propagation between interacting neural populations.  
Additionally, several ``stochastic 
machines'' \cite{haykin:1999} have been formulated, including Boltzmann 
machines \cite{hinton/sejnowski:1986}, sigmoid belief networks 
\cite{neal:1992}, and Helmholtz machines \cite{dayan/hinton:1996}.  Stochastic 
machines are built of stochastic neurons that occupy  one of two 
possible states in a probabilistic manner. Learning rules for 
stochastic machines enable such systems to model the underlying probability 
distribution of a given data set.

The putative modular nature of cortical processing fits well in such a 
probabilistic framework. Cortical areas collectively represent the joint PDF 
over several variables. These neural ``problem-solving modules'' 
can be mapped 
in a relatively direct fashion onto the nodes of a Bayesian belief network,
giving rise to a class of neural network network models that we have termed 
\emph{neural belief networks} 
\cite{barber/clark/anderson:2003a,barber/clark/anderson:2003b}.

In contrast, recent work based on population-temporal coding 
\cite{eliasmith/anderson:1999,eliasmith/anderson:2002} 
indicates that the modeling of low-level 
sensory processing and output motor control do not require such 
a sophisticated representation: manipulation of mean values instead of
PDFs is generally sufficient. Further, the representations can be simplified 
to deal with vector spaces describing the mean values instead of
function spaces describing the probability density functions.

In this work, we develop neural networks processing mean values of analog 
variables as a specialized form of the more general neural belief networks. We 
begin with a brief summary of the key relevant properties of Bayesian belief 
networks in \sectionref{sec:bbns}. We describe a procedure for generating and 
evaluating the neural networks in \sectionref{sec:mvbns}, and apply the 
procedure to several examples in \sectionref{sec:applications}. 

\section{Bayesian Belief Networks}\label{sec:bbns}

\begin{figure}[p]
\centering
\includegraphics[scale=.5]{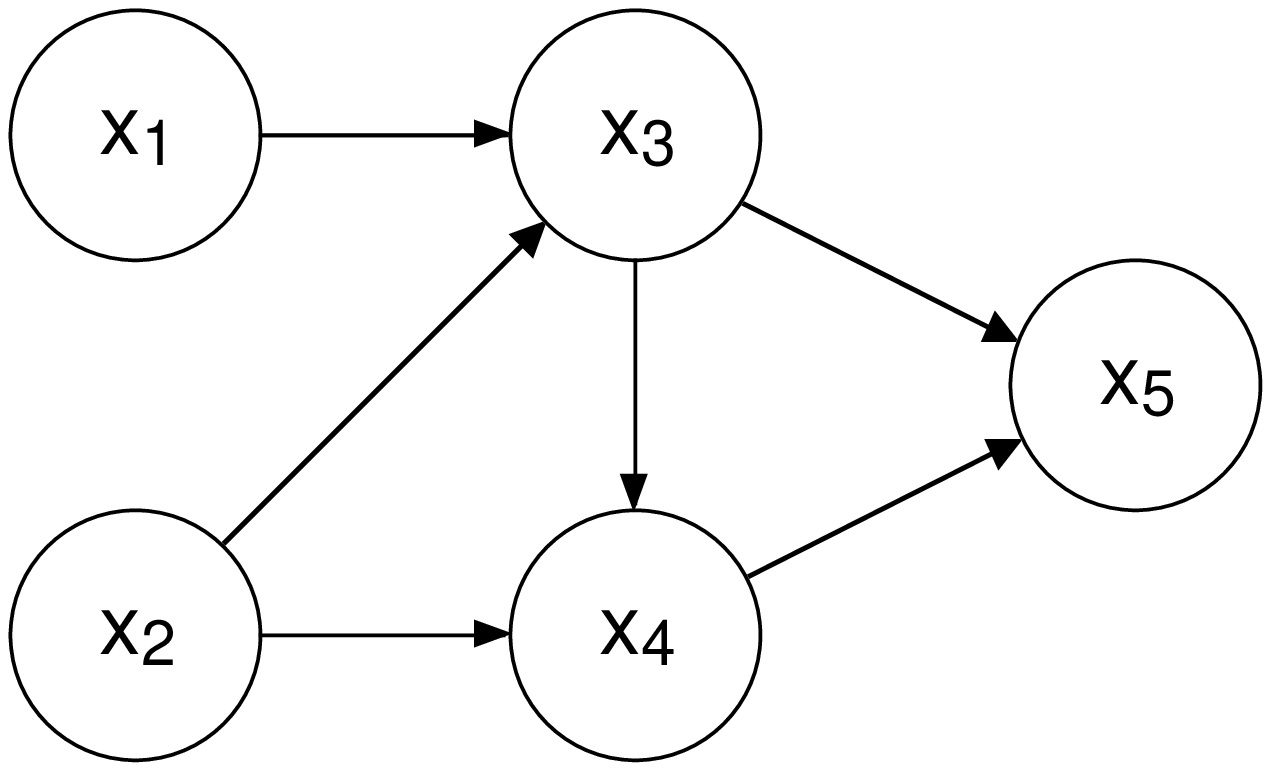}
\caption{A Bayesian belief network.  Evidence about any of the
random variables influences the likelihood of, or belief in, the 
remaining random variables. 
In a straightforward terminology, the node at the tail 
of an arrow is a parent of the child node at the head of the arrow, e.g.
\(X_4\) is a parent of \(X_5\) and a child of both \(X_2\) and
\(X_3\).
From the structure of the graph, we can see the conditional independence 
relations in the probabilistic model.  For example, \(X_5\) is 
independent of \(X_1\) and \(X_2\) given \(X_3\) and \(X_4\).  
}
\label{fig:bbn}
\end{figure}

Bayesian belief networks \cite{pearl:1988,smyth/etal:1997} are directed 
acyclic 
graphs that represent 
probabilistic models (\figref{fig:bbn}).  
Each node represents 
a random variable, and the arcs signify the presence of direct causal 
influences between the linked variables.  The strengths of these 
influences are defined using conditional probabilities.  The 
direction of a particular link indicates the direction of causality (or, 
more simply, relevance); an arc points from cause to effect.

Multiple sources of evidence about the random variables are conveniently
handled using BBNs. The belief, or degree of confidence, in particular 
values of the random
variables is determined as the likelihood of the value given evidentiary
support provided to the network.
There are two types of support that arise from the
evidence: predictive support, which propagates from cause to effect 
along the direction of the arc, and retrospective support, which
propagates from effect to cause, opposite to the direction of the
arc. 

Bayesian belief networks have two properties that we will find very useful, 
both of which stem 
from the dependence relations shown by the graph structure.  First, the value 
of a node $X$ is not dependent upon all of the other graph nodes.  
Rather, it depends only on a subset of the nodes, called a Markov 
blanket of $X$, that separates node $X$ from all the other nodes in 
the graph.  The Markov blanket of interest to us is readily determined from 
the graph structure.  It is comprised of the union of the direct parents 
of $X$, the 
direct successors of $X$, and all direct parents of the direct 
successors of $X$.  Second, the joint probability over the random 
variables is decomposable as
\begin{equation}
    P(x_{1},x_{2},\ldots,x_{n}) = \prod_{\mu=1}^{n} P(x_{\mu}\given 
    \pa{x_{\mu}}) \mathcomma
    \label{eq:bbndecompose}
\end{equation}
where \pa{x_{\mu}} denotes the (possibly empty) set of direct-parent nodes 
of $X_{\mu}$.  
This decomposition comes about from repeated application of Bayes' 
rule and from the structure of the graph.  

\section{Mean-Value Neural Belief Networks}\label{sec:mvbns}

We will develop  neural networks from
the set of marginal distributions \set{\rho(x_{\mu};t)}
so as to best match a desired probabilistic model \( \rho(x_1, x_2,
\ldots, x_D) \) over the set of random variables, which are
organized as a BBN.  One or more of the variables \( x_{\mu} \)  must be
specified as evidence in the BBN. To facilitate the development of general
update rules, we do not distinguish between evidence and 
non-evidence nodes in our notation.  

Our general approach will be to minimize the difference between 
a probabilistic model  \( \rho(x_1, x_2, \ldots, x_D) \) and an estimate 
of the probabilistic model \( \rhoest(x_1, x_2, \ldots, x_D) \). For 
the estimate, we utilize 
\begin{equation}
    \rhoest(x_1, x_2, \ldots, x_D) = 
    \product_{\alpha} \rho(x_{\alpha};t) 
    \mathperiod
    \label{eq:estnaive}
\end{equation}
This is a  so-called naive estimate, wherein the random variables are 
assumed to be independent. We will place further constraints on the 
probabilistic model and representation to produce neural networks with 
the desired dynamics.  

The first assumption we make is that the populations of neurons only
need to accurately encode the mean values of the random variables,
rather than the complete PDFs.  We take the firing rates of the
neurons representing a given random variable \(X_{\alpha}\) to be 
functions of the mean value \meanval{\alpha}(t)
(\figref{fig:firingrates})
\begin{equation}
    \act{\alpha}{i}(t) = \mapfunc{\maplinear{\alpha}{i}
    \meanval{\alpha}(t) + \mapconst{\alpha}{i}}
    \mathcomma 
    \label{eq:neurresponses}
\end{equation}
where 
\maplinear{\alpha}{i}
and 
\mapconst{\alpha}{i} 
are parameters describing 
the response properties of neuron \( i \) of the population
representing random variable \( X_{\alpha} \).  The activation function \(g\)
is in general nonlinear; in this work, we take \(g\) to be the logistic
function,
\begin{equation}
    \mapfunc{x} = \frac{1}{1+\exp\left(-x\right)}
    \mathperiod
\end{equation}
We can make use 
of~(\ref{eq:neurresponses}) to directly encode mean values into 
neural activation states, providing a means to specify the value of the
evidence nodes in the NBN. 

Using~(\ref{eq:neurresponses}), we  derive an update rule
describing the neuronal dynamics,
obtaining (to first order in \(\tau\))
\begin{equation}
    \act{\alpha}{i}(t +\tau) = 
    \mapfunc{\maplinear{\alpha}{i} \meanval{\alpha}(t)
    +\tau \maplinear{\alpha}{i}  \frac{d\meanval{\alpha}(t)}{dt}  +
    \mapconst{\alpha}{i} }
    \mathperiod
    \label{eq:updaterule}
\end{equation}
Thus, if we can determine how \meanval{\alpha} changes with time, we can
directly determine how the neural activation states change with time.

The mean value \(\meanval{\alpha}(t)\) can be determined from the firing
rates as the expectation value of the random variable \(X_{\alpha}\) with
respect to a PDF \(\rho(x_{\alpha}; t)\) represented in terms of some
decoding functions 
\set{\decoder{\alpha}{i}}
The PDF is recovered using
the relation
\begin{equation}
	\rho(x_{\alpha};t) = \sum_i \act{\alpha}{i} \decoder{\alpha}{i}
    \mathperiod
    \label{eq:decodingrule}
\end{equation}
The decoding functions are constructed so as to minimize the difference
between the assumed and reconstructed PDFs (discussed in detail in 
\cite{barber/clark/anderson:2003a}).

With representations as given in %(\ref{eq:updaterule}) and
(\ref{eq:decodingrule}), we have
\begin{eqnarray}
    \meanval{\alpha}(t) & = & \int x_\alpha \rho\left(x_{\alpha}; t\right)
    \, dx_{\alpha} \nonumber \\
    & = & \sum_i \act{\alpha}{i}(t) \decodeconst{\alpha}{i} 
    \mathcomma 
	\label{eq:expectvalues}
\end{eqnarray}
where we have defined
\begin{equation}
    \decodeconst{\alpha}{i} = \int x_{\alpha} \decoder{\alpha}{i}\, dx_{\mu} 
    \mathperiod
\end{equation}
Although we used the decoding functions \(\decoder{\alpha}{i}\) to 
calculate the 
parameters \(\decodeconst{\alpha}{i}\), they can in practice be found directly 
so that the
relations in (\ref{eq:neurresponses}) and (\ref{eq:expectvalues}) are mutually 
consistent.

We take the PDFs \(\rho(x_{\alpha}; t)\) to be normally distributed
with the form
\( \rho(x_{\alpha};t) \equiv \rho(x_{\alpha}; \meanval{\alpha}(t)) 
= N(x_{\alpha}; \meanval{\alpha}(t),
\sigma_{x_{\alpha}}^2)\).  Intuitively, we might expect that the variance
\( \sigma_{x_{\alpha}}^2\) should be small so that the mean value is coded
precisely, but we will see that the variances have no significance in
the resulting neural networks.

The second assumption we make 
is that interactions between the
nodes are linear:
\begin{equation}
    x_{\beta} = \sum_{\alpha}
	\coupling{\beta}{\alpha}
	x_{\alpha} 
    \mathperiod
\end{equation}
Utilizing the causality relations given by the Bayesian belief network,
we require that  \(\coupling{\beta}{\alpha} \neq 0\) only if \(X_{\beta}\) is 
a child node of \(X_{\alpha}\) in the network graph.
To represent the linear interactions as a probabilistic model, we
take the normal distributions \(\rho(x_{\beta} \given \pa{x_{\beta}}) = 
N(x_{\beta};  \sum_{\alpha} 
\coupling{\beta}{\alpha}x_{\alpha}, \sigma_{\beta}^2 ) \) for the
conditional probabilities.

For nodes in the BBN which have no parents, the conditional
probability \(\rho(x_{\beta} \given \pa{x_{\beta}})\) is just the
prior probability distribution \(\rho(x_{\beta})\). We utilize the 
same rule to define the prior probabilities  as to define 
the conditional probabilities. For parentless nodes,
the prior is thus normally distributed with 
zero mean, \(\rho(x_{\beta}) = N(x_{\beta}; 0 , \sigma_{\beta}^2 )\).

We use the relative entropy \cite{papoulis:1991} as a measure of the
``distance'' between the joint distribution
describing the probabilistic model \( \rho(x_1, x_2,
\ldots, x_D)\) and the 
PDF estimated from the neural 
network \(\rhoest(x_1, x_2, \ldots, x_D)\).
Thus, we minimize
\begin{equation}
    E = - \int \rhoest(x_1, x_2, \ldots,x_D) \log \left (
	\frac
	{\rho(x_1, x_2, \ldots,x_D)} 
    {\rhoest(x_1, x_2, \ldots,x_D)}
    \right ) \,
    dx_1 dx_2\cdots dx_D
\end{equation}
with respect to the mean values \meanval{\alpha}.  By making use of the
gradient descent prescription 
\begin{equation}
    \frac{d\meanval{\gamma}}{dt} = -\eta \frac{\partial E}{\partial \meanval{
\gamma}}
\end{equation}
and the decomposition property for BBNs given by (\ref{eq:bbndecompose}),
we obtain the update rule for the mean
values,
\begin{eqnarray}
    \frac{d\meanval{\gamma}}{dt} & = & \frac{\eta}{\sigma_{\gamma}^2} 
    \left( 
    \sum_{\beta}
    \coupling{\gamma}{\beta}\meanval{\beta} - \meanval{\gamma}\right) 
    \nonumber \\
    && -\eta \sum_{\beta}
    \frac{\coupling{\beta}{\gamma}}{\sigma_{\beta}^2}
    \left ( \sum_{\alpha} 
    \coupling{\beta}{\alpha}\meanval{\alpha} - \meanval{\beta} \right )
    \mathperiod
    \label{eq:finalupdate}
\end{eqnarray}
Because the coupling parameters \coupling{\alpha}{\beta} are nonzero 
only when \(X_{\alpha}\) is a parent of \(X_{\beta}\), generally only a 
subset of the mean values contributes to updating \(\xbar_{\gamma}\)
in~(\ref{eq:finalupdate}). In terms of the 
belief network graph structure, the only contributing values 
come from the parents 
of \(X_{\gamma}\), the children of  \(X_{\gamma}\), and the parents 
of the children of \(X_{\gamma}\); this is identical to the Markov blanket
discussed in \sectionref{sec:bbns}.

The update rule for the neural activities is obtained by combining
(\ref{eq:updaterule}), (\ref{eq:expectvalues}), and 
 (\ref{eq:finalupdate}), resulting in
\begin{equation}
    \act{\gamma}{i}(t+\tau) = \mapfunc{ 
        \sum_j \stablewts{\gamma}{i}{j} \act{\gamma}{j}(t) +
        \mapconst{\gamma}{i}+ \eta \tau \externsignal{\gamma}{i}(t)}
		\mathperiod
\end{equation}
The quantity \(\sum_j \stablewts{\gamma}{i}{j} \act{\gamma}{j}(t) +
\mapconst{\gamma}{i}\) serves to stabilize the activities of the neurons 
representing \(\rho(x_{\gamma})\) (similar to neural integrator 
models 
\cite{barber/clark/anderson:2003a,eliasmith/anderson:2002,seung:1996}),
while  
%\externsignal{\gamma}{i}(t)  It is defined as
\begin{eqnarray}
    \externsignal{\gamma}{i} (t) & = &
    \sum_j \selfwts{\gamma}{i}{j} \act{\gamma}{j}(t) \nonumber \\
    &&{} +
    \sum_\beta \sum_j \left(\ffwts{\gamma}{\beta}{i}{j} +
    \fbwts{\gamma}{\beta}{i}{j}\right)\act{\beta}{j}(t) \nonumber \\
    &&{} +
    \sum_{\alpha,\beta} \sum_j 
    \indirectwts{\gamma}{\beta}{\alpha}{j}{i} \act{\alpha}{j}(t) 
%    \mathperiod        
\end{eqnarray}
drives changes in \(\act{\gamma}{i}(t)\)
based on the PDFs represented by other nodes of the BBN.
The synaptic weights of the neural network are
\begin{eqnarray}
    \stablewts{\gamma}{i}{j} & = & 
        \wtdefwrap{\gamma}{i}{j}{} \mathcomma\\
    \selfwts{\gamma}{i}{j} & = & 
        -\wtdefwrap{\gamma}{i}{j}{\varinv{\gamma}} \mathcomma\\
    \ffwts{\gamma}{\beta}{i}{j} & = & 
        \wtdefwrap{\gamma}{i}{j}{\couplevar{\gamma}{\beta}}  \mathcomma\\
    \fbwts{\gamma}{\beta}{i}{j} & = & 
        -\wtdefwrap{\gamma}{i}{j}{\couplevar{\beta}{\gamma}} \mathcomma\\
    \indirectwts{\gamma}{\beta}{\alpha}{j}{i} & = & 
        \wtdefwrap{\gamma}{i}{j}
		{\couplevar{\gamma}{\beta}\coupling{\beta}{\alpha}}
\end{eqnarray}

The foregoing provides an algorithm for generating and evaluating 
neural networks that process mean values of random variables.  To 
summarize,
\begin{enumerate}
    \item  Establish independence relations between model variables.  
    This may be accomplished by  using a graph to organize the 
    variables.
    \item  Specify the \coupling{\alpha}{\beta} to quantify the relations 
between 
    the variables.
    \item  Assign network inputs by encoding desired values into 
    neural activities using~(\ref{eq:neurresponses}).
    \item  Update other neural activities
    using   (\ref{eq:finalupdate}).
    \item  Extract the expectation values of the variables from the neural 
    activities using (\ref{eq:expectvalues}).
\end{enumerate}

\section{Applications}\label{sec:applications}

\begin{figure}[p]
\centering
\includegraphics[width=3.25in]{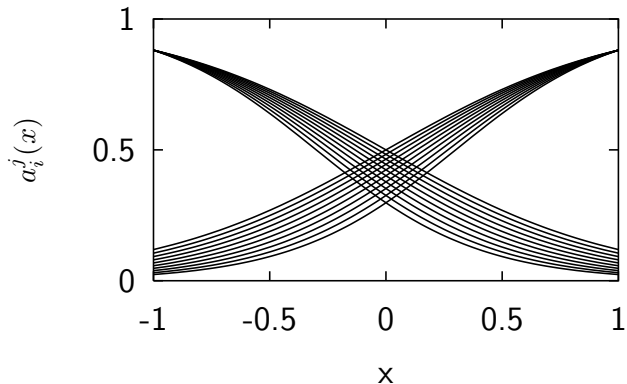}
\caption{The mean values of the random variables are encoded into
the firing rates of populations of neurons.  A population of twenty
neurons with piecewise-linear responses is associated with each 
random variable.  The neuronal responses \act{\alpha}{i} are fully determined by 
a
single input \( \xi \), which we interpret as the mean value of a
PDF.  
The form of the neuronal transfer functions can be altered without
affecting the general result presented in this work.
}
\label{fig:firingrates}
\end{figure}

As a first example, we apply the algorithm to the BBN shown in 
\figref{fig:bbn}, with firing rate profiles as shown in
\figref{fig:firingrates}. Specifying \(x_1 = 1/2\) and \(x_2 =
-1/2\) as evidence, we find an excellent match between the mean values
calculated by the neural network and the directly calculated values
for the remaining nodes (Table~\ref{table:comparison}).

\begin{figure}[p]
    \centering
    \subfigure[]{\includegraphics[scale=.5]{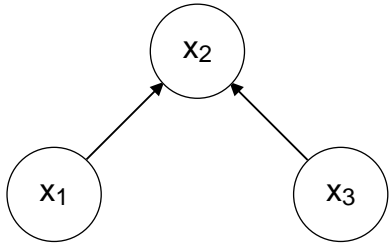}}
    \hfil
    \subfigure[]{\includegraphics[scale=.5]{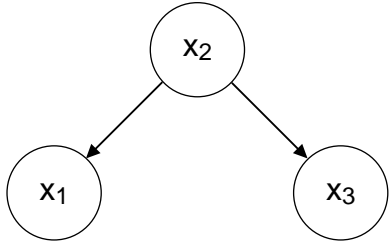}}
    \caption{Simpler BBNs.  Although the underlying undirected graph 
    structure is identical for these two networks, the direction of 
    the causality relationships between the variables are reversed.   
    The neural networks arising from the BBNs thus have different 
    properties.}
    \label{fig:trees}
\end{figure}

We next focus on some simpler BBNs to highlight certain 
properties of the resulting neural networks (which will again utilize 
the firing rate profiles shown in 
\figref{fig:firingrates}).  In 
\figref{fig:trees}, we present two BBNs that relate three random 
variables in different ways.  The connection strengths are all taken 
to be unity in each graph, so that \( \coupling{2}{1} = \coupling{2}{3} 
= \coupling{1}{2} = 
\coupling{1}{3} = 1\).  

With the connection strengths so chosen, the two BBNs have  
straightforward interpretations.  For the graph shown in 
\figref{fig:trees}a, \( X_{2} \) represents the sum of \( X_{1} \) 
and \( X_{3} \), while, for the graph shown in 
\figref{fig:trees}b, \( X_{2} \) provides a value which is 
duplicated in \( X_{1} \) and \( X_{3} \).  The different graph 
structures yield different neural networks; in particular, nodes 
\( X_{1} \) and \( X_{3} \) have direct synaptic connections in the 
neural network based 
on the graph in 
\figref{fig:trees}a, but no such direct weights exist in a second 
network based on
\figref{fig:trees}b.  Thus, specifying \( x_{1} = -1/4\) and 
\( x_{2} = 1/4 \) for the first network produces the expected result 
\( \xbar_{3} = -0.5000 \), but specifying \( x_{2} = 1/4 \) in the second 
network produces  
\( \xbar_{3} = 0.2500 \) regardless of the value (if any) 
assigned to \( x_{1} \).   

To further illustrate the neural network properties, we use
the graph shown in \figref{fig:trees}b to process 
inconsistent evidence.  Nodes \( X_{1} \) and \( X_{3} \) should 
copy the value in node \( X_{2} \), but we can specify any values we 
like as network inputs.  For example, when we assign \( x_{1} = -1/4 \) and 
\( x_{3} = 1/2 \), the neural network yields \( \xbar_{2} = 0.1250 \) for 
the remaining value.  
This is a typical and reasonable result, matching
the least-squares solution to the inconsistent problem.

\section{Conclusion}\label{sec:concl}

We have introduced a class of neural networks that
consistently mix multiple sources of evidence. The networks are
based on probabilistic models, represented in the graphical form of
Bayesian belief networks, and function based on traditional
neural network dynamics (\ie, a weighted sum of neural 
activation values passed through a nonlinear activation function).
We constructed the networks by
restricting the represented probabilistic models by introducing 
two auxiliary assumptions.  

First, we assumed that only the mean values of the
random variables need to be accurately represented, with higher
order moments of the distribution being unimportant. We introduced
neural representations of relevant probability density functions
consistent with this assumption.  Second, we assumed that the random
variables of the probabilistic model are linearly related to one another, 
and chose appropriate
conditional probabilities to implement these linear relationships. 

Using the representations suggested by our auxiliary assumptions, we
derived a set of update rules by minimizing the relative entropy
of an assumed PDF with respect to the PDF decoded from the neural
network.  In a straightforward fashion, the optimization procedure
yields neural weights and  dynamics that implement specified probabilistic 
relations, without the need for a training process.

The restricted class of neural belief networks investigated in this work
captures many of the
properties of both Bayesian belief networks and neural networks.  
In particular, 
multiple sources of evidence are consistently pooled based on local
update rules, providing a distributed version of a
probabilistic model. 

%: Acknowledgments
\begin{acknowledgments}
    This work was supported in part by the U. S. National Science 
    Foundation under Grant PHY-0140316 and in part by the Portuguese 
    Funda{\c c}\~ao para a Ci\^encia e a Technologia (FCT) under Bolsa
    SFRH/BPD/9417/2002). JWC also acknowledges support received from 
    the Funda{\c c}\~ao Luso-Americana para o Desenvolvimento (FLAD)
    and from the FCT for his participation in Madeira Math Encounters XXIII at
    the University of Madiera, where portions of the work were conducted.
\end{acknowledgments}

\begin{table}[p]
    \centering
    \caption{The mean values decoded from the neural network closely
    match the values directly calculated from the linear relations.  The
    coefficients for the linear combinations were randomly selected,
    with values \( \coupling{3}{1} = -0.2163\), \(\coupling{3}{2} = -0.8328\), 
    \(\coupling{4}{2} =
    0.0627\), \(\coupling{4}{3} = 0.1438\), \(\coupling{5}{3} = -0.5732\), 
    and \(\coupling{5}{4} =
    0.5955\). } \label{table:comparison}
    \begin{tabular}{c|c|c}
	Node & Direct Calculation & Neural Network \\ \hline
	\(X_1\) &\ 0.5000 &\ 0.5000 \\
	\(X_2\) & -0.5000 & -0.5000\\
	\(X_3\) & \ 0.3083 & \ 0.3084\\
	\(X_4\) & \ 0.0130 & \ 0.0128\\
	\(X_5\) & -0.1690 & -0.1689
    \end{tabular}
\end{table}

\end{document}